# Scattering-Induced Mode Chirality in Ring Resonators

**Authors**
Haochen Yan[1, 2†], Xu Guo[3†], Arghadeep Pal[1, 2†], Xiaoyuan Huang[4, 5†], Alekhya Ghosh[1, 2†], Lewis Hill[1], Shuangyou Zhang[1, 6], Nivedita Vishnukumar[1, 2], Toby Bi[1, 2], Masoud Kheyri[1, 2], Jianming Mai[7], Hao Zhang[1], Yaojing Zhang[8], Jolly Xavier[9], Haihua Fan[3], Kok Wai Cheah[7*], Peter Littlewood[5, 10, 11*], Pascal Del'Haye[1, 2*]

**Affiliations**
[1]*Max Planck Institute for the Science of Light, Staudtstr. 2, 91058, Erlangen, Germany*
[2]*Department of Physics, Friedrich Alexander University Erlangen-Nuremberg, 91058, Germany*
[3]*Guangdong Provincial Key Laboratory of Nanophotonic Functional Materials and Devices, School of Optoelectronic Science and Engineering, Guangdong Basic Research Center of Excellence for Structure and Fundamental Interactions of Matter, South China Normal University, Guang-Zhou 510006, China*
[4]*Pritzker School of Molecular Engineering, University of Chicago, Chicago, IL60637, USA*
[5]*James Franck Institute, University of Chicago, Chicago, IL60637, USA*
[6]*Department of Electrical and Photonics Engineering, Technical University of Denmark, Kgs. Lyngby 2800, Denmark*
[7]*Department of Physics, Hong Kong Baptist University, Kowloon Tong, Hong Kong SAR, China*
[8]*School of Science and Engineering, The Chinese University of Hong Kong, Shenzhen, Guangdong 518172, China*
[9]*SeNSE, Indian Institute of Technology Delhi, Hauz Khas, 110016, New Delhi, India*
[10]*Department of Physics, University of Chicago, Chicago, IL60637, USA*
[11]*School of Physics and Astronomy, University of St Andrews, St Andrews, KY16 9AJ, UK*

[†]These authors contributed equally to this work
[*] Corresponding authors

## Abstract
Non-Hermitian physics can be used to break time reversal symmetry and is important for interactions in a wide range of systems, from active matter and neural networks to metamaterials and non-equilibrium thermodynamics. In integrated photonic devices, non-Hermitian physics can be used for direction-dependent light propagation, reconfigurable light paths, selective energy localization and optical isolators. In this work, we report previously unexplored direction-dependent mode splitting in ring microresonators, achieved by adding multiple scatterers around the cavity. Through experiments, simulations, and theoretical modeling, we unveil the underlying physics that changes the resonance shapes in resonant systems with backscattering. By engineering the spatial configuration of the scatterers, we can produce a predictable and repeatable direction-dependent mode splitting, enabling new ways to route light through optical resonators and photonic networks. In addition, the direction dependent mode-splitting can be used for precise near-field measurements, enhancing traditional sensing in integrated photonic chips.

# Main

Non-Hermitian physics, defined by breaking the time-reversal symmetry[1-3], underpins a wide range of phenomena including nonreciprocal light propagation[4], stabilizing dynamic systems[5], and protecting topological states[6]. The underlying physics has implications in diverse areas such as condensed matter physics,[7] quantum information[8,9], and photonics[10-12]. Applications range from spintronic devices[13] to precision sensing[14] and optical communications[15]. Integrated photonic systems are very promising for implementing non-Hermitian physics, since they are compact, efficient, and scalable. In this context, microresonators[16] are of particular interest for exploring non-Hermitian effects[17-19]. They have the ability to confine light within ultra-small volumes and are widely used in integrated photonic circuits. Microresonators have been used to study optical effects such as mode splittings[20], backscattering-induced asymmetries[21-24], and unidirectional light propagation[25], which make them ideal for unraveling fundamental physics[26] and engineering new functionalities[5,14,27,28].

Here, we report a previously unobserved direction-dependent mode splitting in ring microresonators[29] achieved by introducing scatterers[22,30,31]. The experiment is schematically depicted in Fig. 1(a). The splitting is only observed in one direction and the circulating intensities in the resonator vary when changing the scatterers' spatial separation. The splitting can be intuitively explained as arising from non-Hermitian external coupling, which induces asymmetric backscattering and alters the spectral shape of the modes. In addition, the asymmetry of the mode splitting varies when the position of one of the scatterers is changed, which is illustrated in Fig. 1(b). In this article, we comprehensively study this phenomenon based on experimental measurements, simulations, and theoretical analysis, and shed new light on the underlying interaction between cavity modes and asymmetric coupling mechanisms. From the application side, the asymmetric mode splitting enables the microresonator to respond very sensitive to external perturbations. We present proof-of-concept demonstrations of utilizing this for sensing applications. These finding can lead to enhanced sensing of physical properties like temperature, length, refractive index or magnetic fields. In addition, it can be used for efficient evanescent field sensing e.g. for the detection of biomolecules.

# Results

In the experiment, we pump a coupling waveguide from one direction and read both the transmission and reflection signals from the drop-port waveguide (see SI, section I). The observation of asymmetric splitting is shown in Fig. 2(a), where we select four results at different pumping frequencies. The transmission signal is shown as a blue line, while the reflection signal is shown as red line. The selected spectra represent four categories of all spectra: i) Asymmetric unequal-intensity splitting; ii) Asymmetric equal-intensity splitting; iii) Weak asymmetric unequal-intensity splitting; and iv) No-splitting as the reference. We refer to the asymmetric splitting as mode splitting that is only observed in one direction. While previous studies have investigated splitting phenomena occurring in both the transmission and reflection spectra (iii-like case) by positioning scatterers near the microresonator, the asymmetric splitting that occurs exclusively in one direction (i and ii cases) has not yet been reported. In addition, the line shape of the split resonances resembles a Fano line shape[32,33], where a sharp drop happens at the splitting. This is also different from previous studies on chiral physics in microresonators. The appearance of the splitting indicates the intrinsic symmetry of the system is broken by introducing the scatterers around the ring microresonator. To demonstrate the broadband working range of this design, we scan the input laser frequency across a 14-THz-window starting from 184 THz to 198 THz. Fig. 2(b) depicts the observed splitting in the different resonances within this frequency window. It is noteworthy that the splitting width varies with frequency and disappears in some regions, confirming the theoretical predictions. The data in Fig. 2(a) shows a selection of different resonance spectra in the two propagation directions.

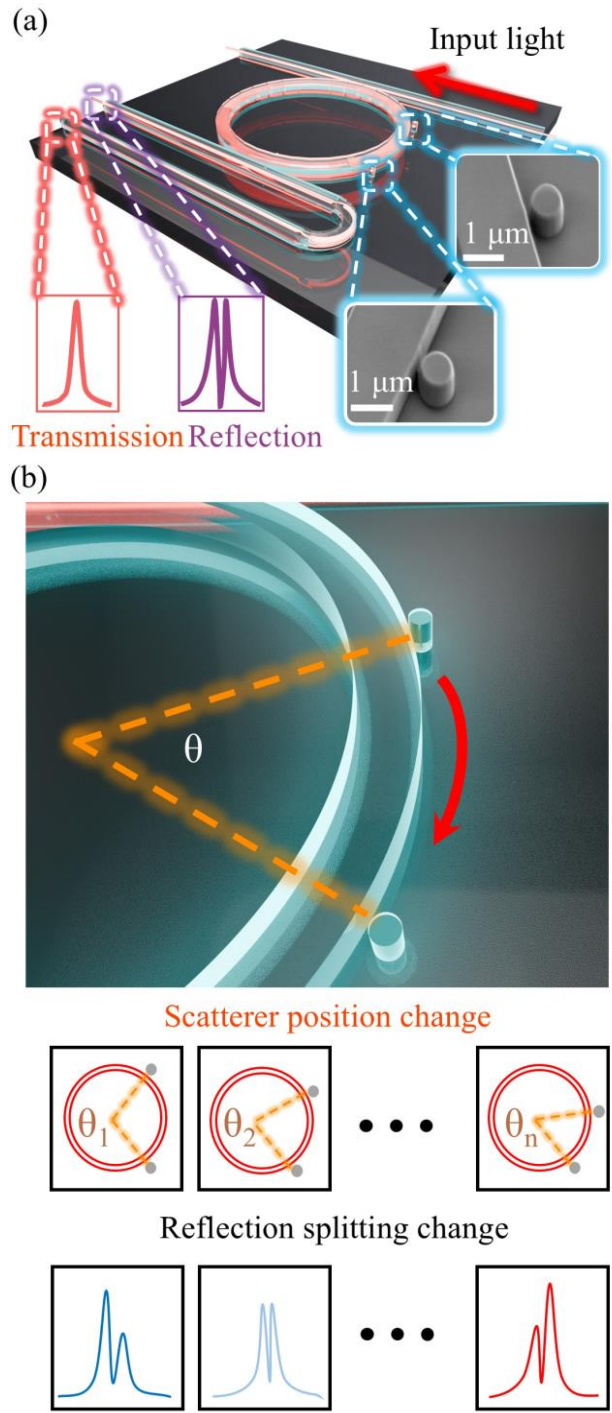

**Figure 1.** Scheme for realizing non-symmetric splitting from a ring-microresonator system with two scatterers. (a) Illustration of the system. The light is coupled into the microresonator via an input waveguide while the transmission and reflection signals are read from a second outcoupling waveguide. Two scatterers are fabricated close to the edge of the microresonator. The insets show scanning electron microscope (SEM) images of the scatterers. A mode splitting is only observed in the reflection direction. (b) Change of the asymmetric mode splitting with different scatterer positions.

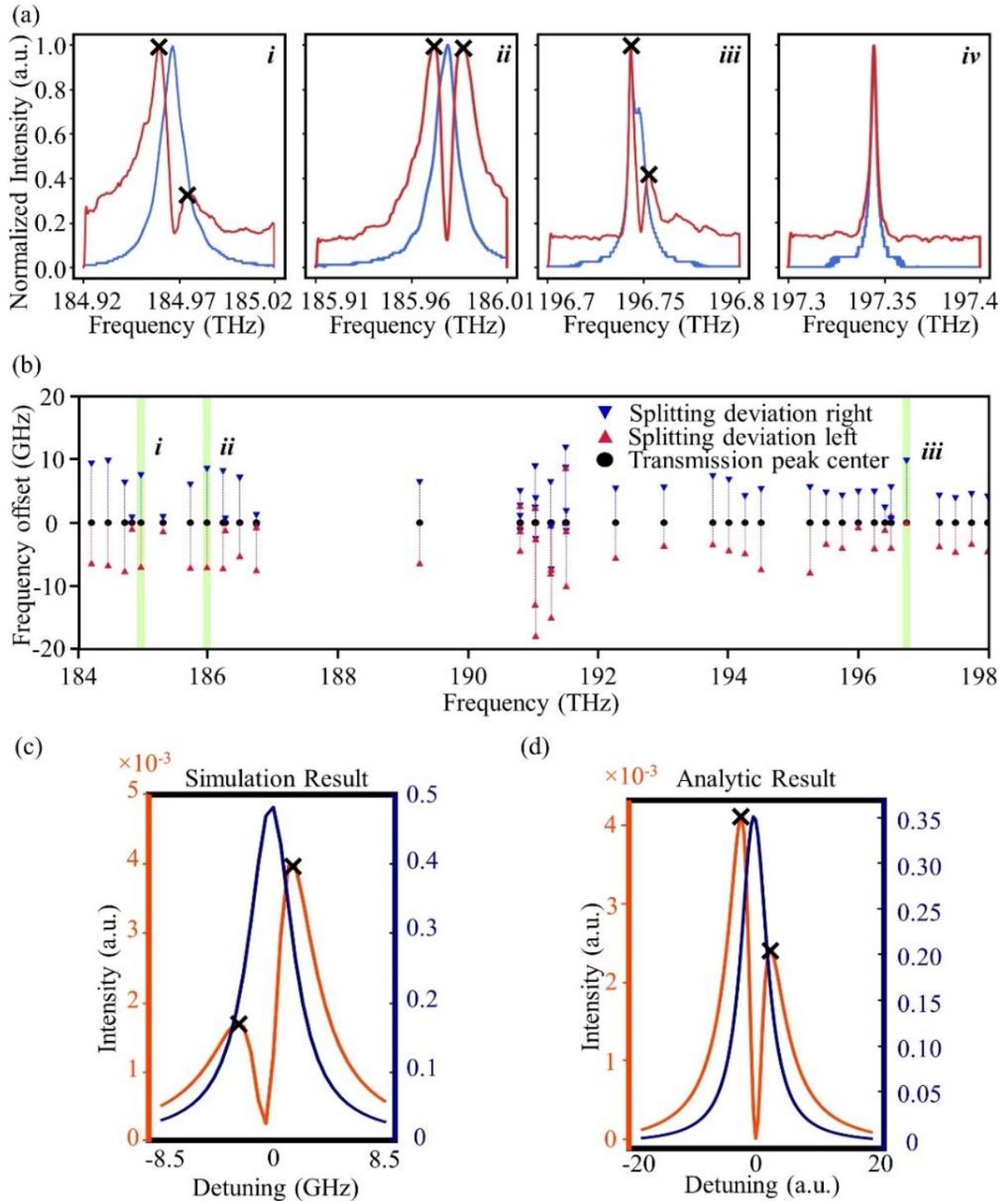

**Figure 2.** Asymmetric mode splitting. (a) Selection of measured resonance spectra with and without mode splitting. *i*. Asymmetric unequal-intensity splitting; *ii*. Asymmetric equal-intensity splitting; *iii*, Asymmetric splitting in both the reflection (strong) and transmission (weak); *iv*. No splitting. The peaks of the split resonances are marked with a black cross. (b) Splitting of different microresonator modes, measured in reflection. Labels *i*-*iii* correspond to the spectra in panel (a). (c) Finite element simulation of reflection (red) and transmission (blue) of a microresonator with two scatterers and asymmetric mode splitting. (d) Analytic result of reflection (red) and transmission (blue) with a modified two-mode-approximation (TMA) model.

To understand and explain the observed mode shapes, we first perform finite element method (FEM) simulations to reproduce the experimental results, as shown in Fig. 2(c), where the orange line indicates the reflection spectra and the navy-blue line indicates the transmission spectra. In order to simplify the calculation, we use a two-dimensional model (see SI, section III). We consider a single-frequency pump, and successfully obtain a Lorentzian-shaped transmission signal and an asymmetric splitting in the reflection signal, which exhibits a distinct "Fano-like" profile. The reflection signal intensity is approximately two orders of magnitude smaller than the transmission signal, and it is suppressed to nearly zero at the transmission frequency, which is consistent with the experimental observations. Furthermore, broadband pumping simulations reveal that the unequal-intensity splitting dominates in the splitting cases, aligning well with the summarized splitting results shown in Fig. 2(b) (see SI, section III). The simulation results are in excellent agreement with the experimental observations, providing a solid foundation to characterize the system.

Building on this, we theoretically model the observed dynamics of asymmetric backscattering in a ring cavity system, where we adapt a mean-field approximation using a coupled mode equations[20,34]. Introducing two external scatters into the ring cavity results in asymmetric coupling between the CCW and CW modes, and thus causes the interaction matrix to become non-diagonalizable. This leads to the two eigenmodes become collinear, corresponding to the coalescence of standing modes within the cavity (see SI, section IV). The governing equations of the system are derived by considering the coupling between the CCW and CW intra-cavity modes, and by assuming a constant input $\alpha_{in}$. We can write the equations as:

$$i\partial_t \begin{pmatrix} \Psi_{ccw} \\ \Psi_{cw} \end{pmatrix} = \begin{pmatrix} \Omega & A \\ B & \Omega \end{pmatrix} \begin{pmatrix} \Psi_{ccw} \\ \Psi_{cw} \end{pmatrix} + i\sqrt{\kappa} \begin{pmatrix} \alpha_{in} \\ 0 \end{pmatrix}, \qquad (1)$$

where $\Omega = \omega_0 - i\gamma + 2\tilde{K}$, $A = \tilde{K} e^{-2im\beta_1} + \tilde{K} e^{-2im\beta_2}$, and $B = \tilde{K} e^{2im\beta_1} + \tilde{K} e^{2im\beta_2}$. The integer number m represents the angular mode number, and $\gamma$ is the effective loss rate. $\tilde{K} = K \cdot e^{i\delta_k}$ is a complex phenomenological parameter that accounts for the shifts in frequency and changes in decay rates due to the presence of external scatterers (the real part of $\tilde{V}$ accounts for the frequency shift, while the imaginary part is associated with the modification of the decay rate), and $\beta_1$ and $\beta_2$ are the corresponding azimuthal angular position of the scatterers. As shown in the equation, the non-Hermitian coupling between CW and CCW modes can be tuned through the position of the scatters. Note that the non-Hermitian nature of the system intrinsically gives rise to EP physics, where the system exhibits critical behavior when the determinant of the coupling matrix, $\sqrt{AB}$, approaches zero. We neglect the backscattering in the input waveguide and we add backscattering to the read-out waveguide, defined by a reflection coefficient $w_b$, which interferes with the light from the cavity modes. The $w_b$-term in the following equation accounts for boundary effects, which may unavoidably arise from the input and/or output waveguides. Considering only the output contribution, we hypothesize that additional reflections contribute to a continuum mode that interferes with the output modes emerging from the cavity. Such a boundary effect is significant only in the reflection direction (since the reflection signal is relatively weak) but is negligible in the transmission direction. As a result, the asymmetric splitting is only observed in the reflection direction. We derive the expression for both the transmission and reflection spectra as:

$$\text{Transmission} = \left| \frac{2i\kappa \cdot \Omega}{4\cos^2(m(\beta_1 - \beta_2)) \cdot \widetilde{K}^2 + (i\Omega - \omega)^2} \right|^2, \qquad (2a)$$

$$\text{Reflection} = \left| \frac{2i \cdot \alpha_{\text{in1}} \cdot \gamma_{\text{ex}} \cdot (2 \cdot e^{i[-2m(\beta_1-\beta_2)+2m\theta_1]}) \cdot \widetilde{K} \cdot \cos(m(\beta_1-\beta_2)) - w_b \cdot e^{im(\beta_1-\beta_2)} \cdot \Omega}{4\cos^2(m(\beta_1-\beta_2)) \cdot \widetilde{K}^2 + (i\Omega - \omega)^2} \right|^2, \qquad (2b)$$

where the additional reflection interferes with the CCW and CW modes, leading to an imbalance in the signal intensities. We plot the above transmission and the reflection at a specific angle separation of the scatterers as shown in Fig. 2(d), using the same color coding used in Fig. 2(c). Here, we can clearly see the Lorentzian-shaped transmission signal and asymmetric splitting in the reflection signal, which is around two-orders of magnitude smaller than the transmission signal and becomes nearly-zero at the transmission center frequency. This indicates that we reach the EP condition because the CW mode is effectively suppressed, resulting in the dominance of the CCW mode. The theoretical results presented here agree with the simulation results shown in Fig. 2(c) as well as with the experimental observations in Fig. 2(a). Thus, we conclude that our model provides an accurate description of the asymmetric response in transmission and reflection.

To further explore the mode splitting, we vary the spatial angle between the two scatterers. The experimental results are shown in Fig. 3, where we fabricate multiple resonators on the same chip with varying spatial angular separation between the scatterers. In this figure the blue lines represent the transmission signal and the red lines represent the reflection signal. Selected transmission and reflection spectra across different devices, for one specific frequency window are shown in Fig. 3(a). Different devices are labelled by the spatial separation angle. The first device is fabricated without scatterers and serves as a reference. The dashed lines represent the reflection signals, while the solid lines represent transmission signals in Fig. 3(a). The addition of the scatterers causes the mode splitting, while the variation of the angle between the scatterers leads to a different shape of the split modes. This agrees with the theoretical model, where the spatial angle between the scatterers determines the output signal's shape. To investigate further, we track the evolution of the splitting in a specific frequency window for three distinct angular separations in Fig. 3(b). We observe a transformation in the asymmetric splitting profile when the spatial angle separation is changed from 47° to 61.2°. Specifically, at 47° the low-frequency splitting peak has higher intensity relative to the high-frequency peak. In contrast, at 61.2°, the splitting asymmetry is reversed. The above observation is also predicted by our analytical model, in which a change of the spatial angle between scatterers influences the phase relationship and interference between the counter-propagating modes. As a result, there is a change in the relative height of the peaks.

Figure 3(c) shows another phenomenon at a different pump frequency. When varying the angle from 49° to 60°, we find that the transmission spectra exhibit an "asymmetric EP"-like shape, where the weak splitting peaks in the transmission vanish and then split again. This is different from an EP, where the splitting vanishes in both transmission and reflection[20]. Our model can predict the disappearance of the splitting in transmission at a specific separation angle when the coupling between counter-propagating modes is perfectly balanced. When the system is in the vicinity of an EP, which is achieved by changing the angular positions of the scatterers, the interference effects due to waveguide backscattering become significant enough, such that the cavity mode in one direction (reflection) is suppressed. This corresponds to the condition where the off-diagonal coupling term, *B*, approaches zero, which happens in the proximity of the EP. These findings show that the observed asymmetric backscattering is a manifestation of the underlying exceptional point physics, with interference from waveguide backscattering playing a key role in shaping the system's response. In addition to these observations, we further investigate how the intrinsic resonator properties affect the splitting behavior. Thus, we investigate the splitting width (defined by the spectral separation of the two splitting peaks) as a function of the quality factor for different angular separation of the scatterers, as shown in Fig. 3(d). It can be observed that the splitting width decreases approximately linearly for increasing Q-factors. This trend is expected, since the splitting needs to happen within the linewidth of the resonances, which decreases with increasing Q-factor.

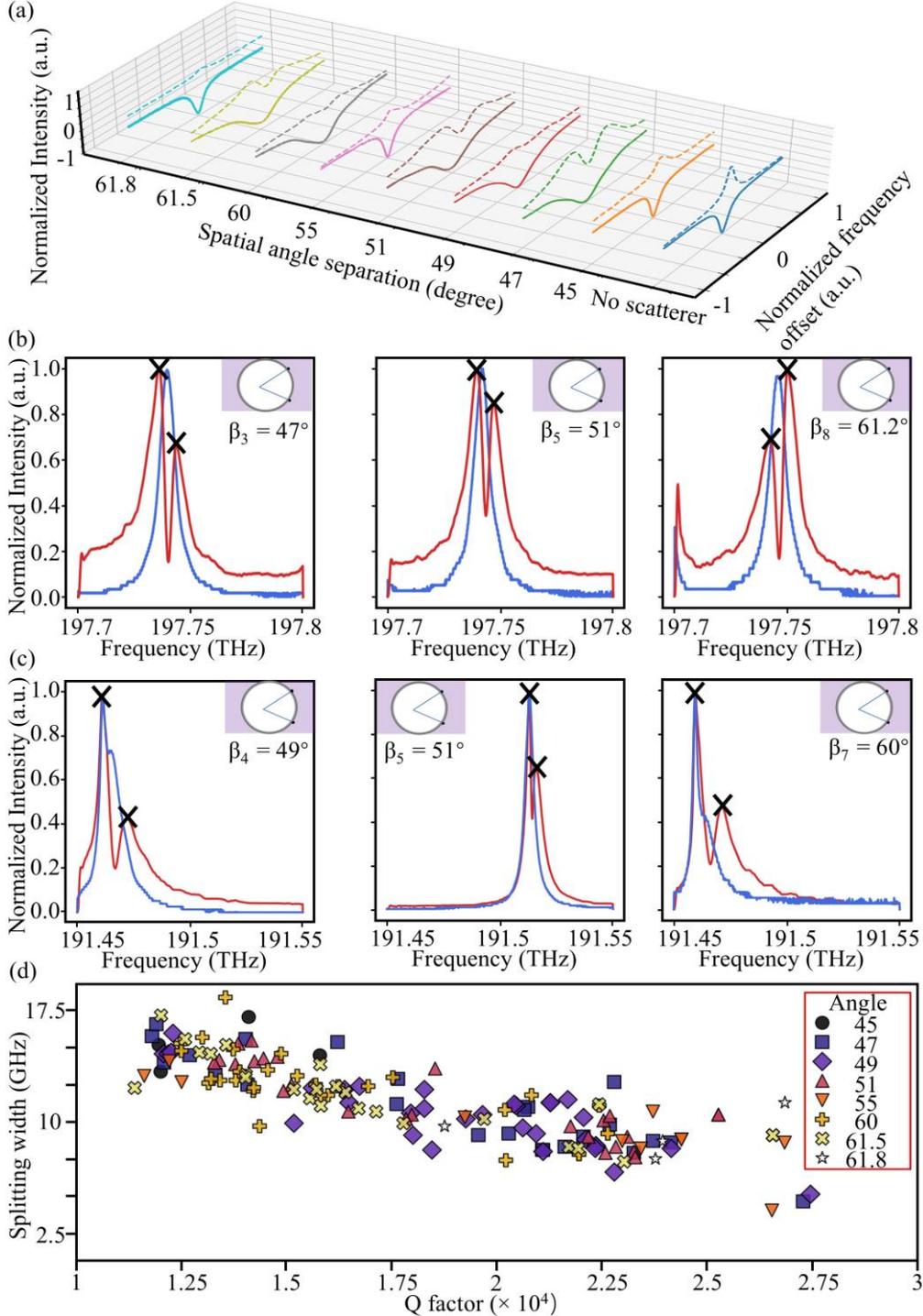

**Figure 3.** Change of asymmetric mode splitting in fabricated devices with varying spatial angle separation between the two scatterers. (a) Evolution of one specific mode in devices with different spatial separations of the scatterers. (b) Splitting change at a fixed optical frequency for different spatial angles between the scatterers. (c) Splitting change at another optical frequency. In this case, the reflection signal remains split, whereas the transmission evolves from weak splitting to no-splitting to weak splitting, which is similar to the behavior around an exceptional point. (d) Splitting widths across all tested devices. The resonance splitting width decreases with increasing Q-factor.

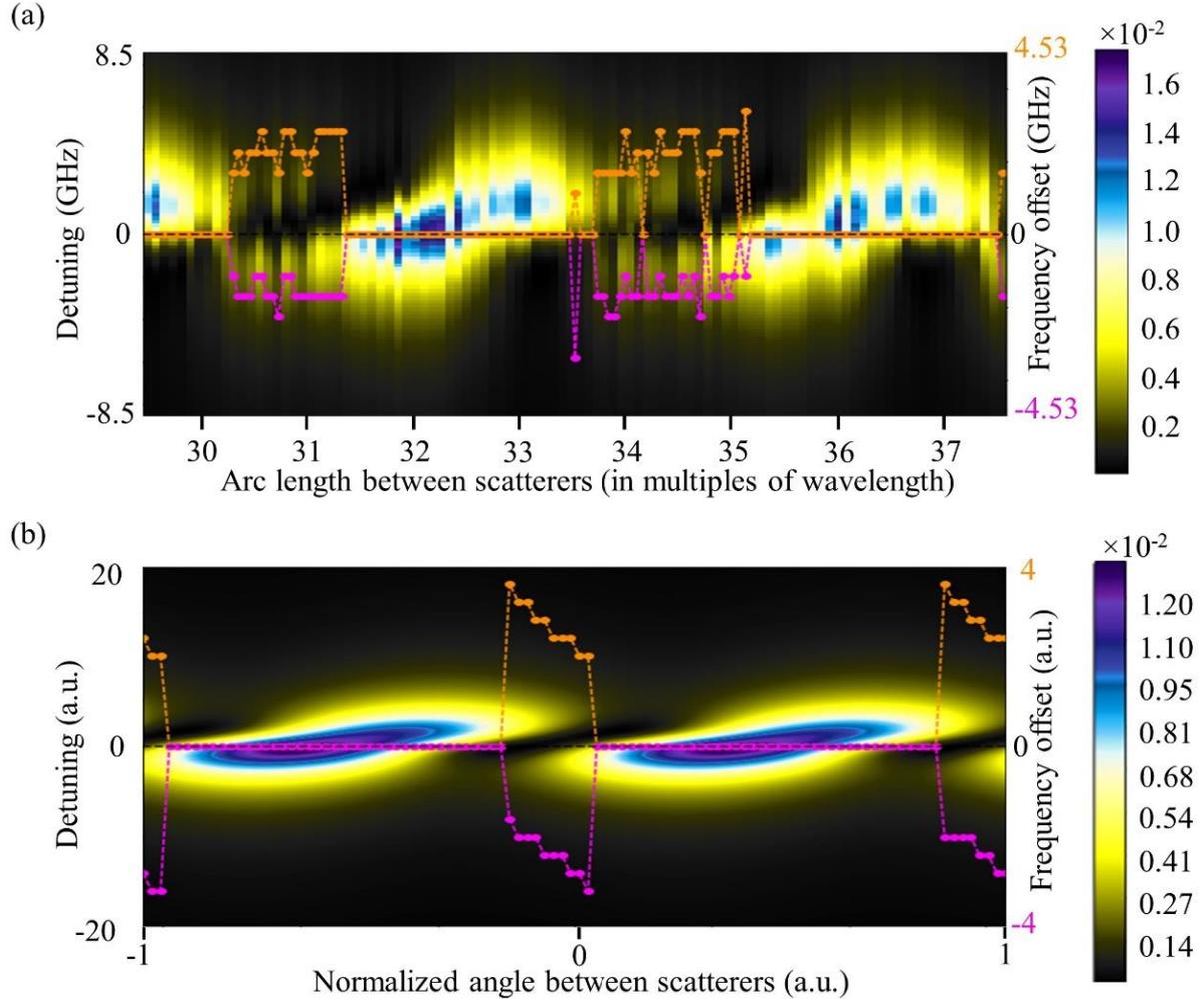

**Figure 4.** Change of mode splitting when changing the angle between scatterers. (a) Finite element simulation results of the mode splitting in reflection when continuously changing the angular separation of the scatterers. The 2D-heat map shows a repeating pattern. The orange and pink lines show the magnitude of the splitting between the peaks. (b) Results from the analytical model for the mode splitting in reflection.

For a more comprehensive understanding of the splitting behavior, we perform a systematic analysis by varying the angle separation continuously. This is done via FEM simulations, and the resulting reflection signals are shown in Fig. 4(a). Here, we show and compare the mode splitting from experimental measurements and theoretical calculations. The reflection spectrum shows a repetitive pattern of regions with splitting and without splitting. Within the parameter ranges that show resonance splitting, the intensity of the reflected light is strongly suppressed. This observation agrees with previous work, in which a tungsten tip can be used to suppress back scattering[21]. In addition, the split resonance peaks are asymmetric in amplitude, which agrees with the experimental results. These simulations confirm that the splitting is highly sensitive to the change in spatial separation of the scatterers. For a more comprehensive analysis, we compare the mode splitting behavior with the results from our analytical model shown in Fig. 4(b), which reproduces the pattern from the numerical simulations. Using the analytical model and numerical simulations, we can engineer the splitting behavior (e.g. periodicity and splitting width) by adjusting the parameters of the model, which can be used to fabricate resonators with specific properties like zero-reflection at specific wavelengths. Additional simulation results are in the supplementary information section VI.

In this last part, we investigate the use of the asymmetric resonance splitting for sensing. Specifically, when a perturbation is introduced, it alters the system, shifting it away from the condition under which an exceptional point (EP) emerges. This measurable change in the reflection spectrum enables the detection of subtle variations in the environment of the resonator with high sensitivity. For sensing, we consider one scatterer in the above analysis as fixed during the chip during fabrication, while the other scatterer can be a free moving particle in the near-field of the resonator. The proof-of-concept results based on the theoretical model are shown in Figure 5. As illustrated in Figs. 5(a) and 5(b), asymmetric splitting appears only in devices with external coupling, i.e. the intra-cavity modes coupled with input/output light. Whereas, without this external coupling, the reflection spectrum remains symmetric and does not exhibit direction-dependent splitting, even in the presence of perturbations. The presence of asymmetric splitting enables highly sensitive detection of weak perturbations, as shown in Fig. 5(c)*i*. When the angular separation between the two scatterers is changed, the reflection spectrum exhibits systematic variations in both the frequency spacing and the relative peak heights of the split modes. These two spectral signatures act at two different scales—similar to the coarse and fine readings on a micrometer screw gauge—providing a precise measurement tool for quantifying small perturbations. Moreover, the response is directional: increasing the angle shifts the peak height asymmetry in one direction, while decreasing it reverses the direction (Fig. 5(c)*ii*). This allows us to not only measure the magnitude but also the direction of a moving perturbation to be measured in real time. Such a perturbation can be caused by any object that introduces localized scattering or loss near the cavity. This makes the scheme widely applicable to a range of sensing scenarios. For comparison, Fig. 5d shows a calculation of a system without external waveguide coupling. In this configuration, the reflection response remains symmetric, and perturbations only cause uniform changes in signal intensity—without peak splitting or directional information—limiting the precision of detection. Furthermore, while the system periodically exhibits spectral coalescence behavior reminiscent of exceptional points when the scatterer angle is changed, our detection method does not rely on the EP itself. Instead, we exploit the asymmetric response induced by the scatterers, enabling continuous and directionally sensitive measurements. The conceptual comparison in Fig. 5 highlights how our approach adds possibilities for more precise measurements, surpassing conventional EP-based sensing schemes that depend solely on modal coalescence or critical loss enhancement.

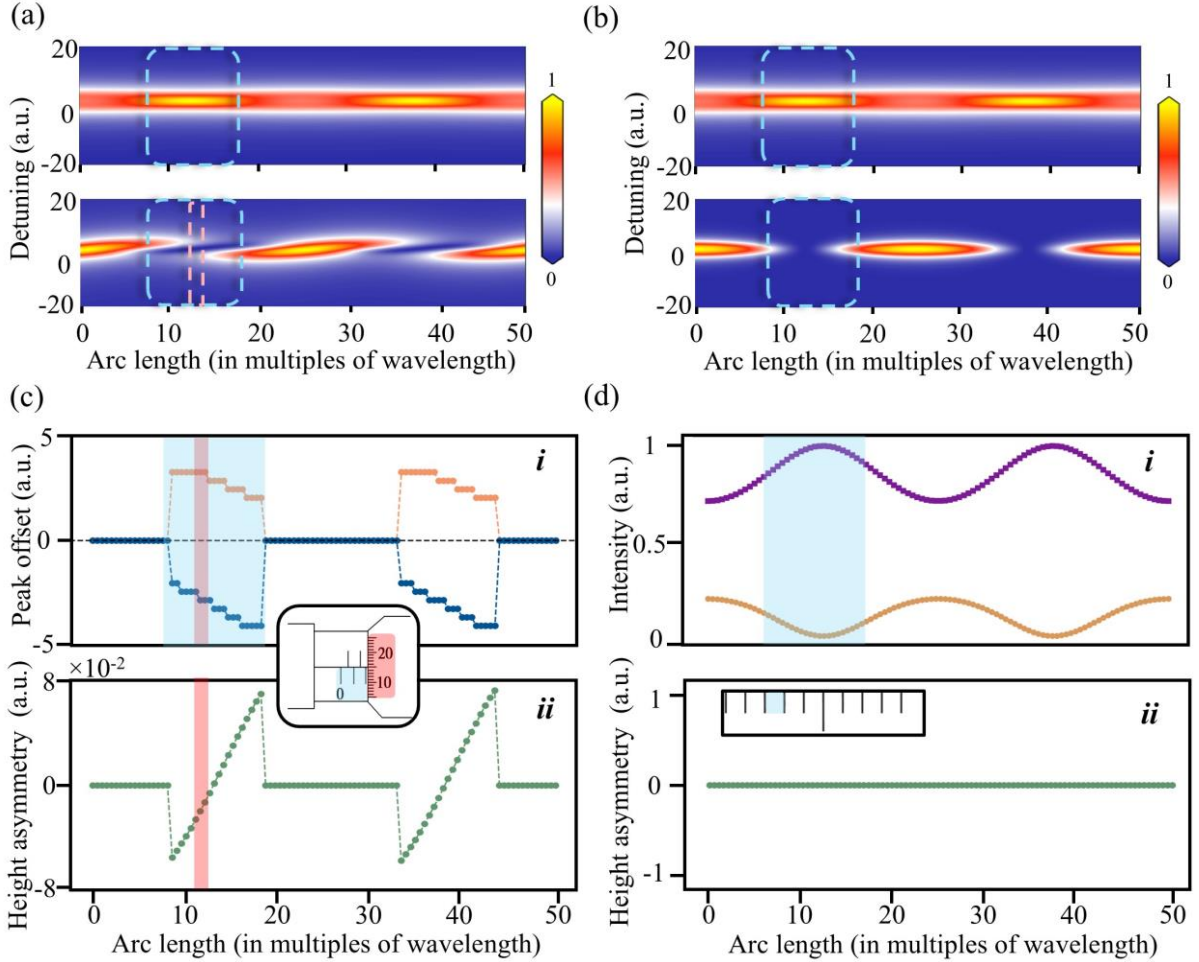

**Figure 5.** Application of the non-symmetric mode splitting as a "micrometer screw". (a) and (b) 2D heatmaps for transmission (top) and reflection (bottom) spectra as functions of detuning and angle between scatterers. (a) shows non-symmetric mode splitting in the reflection and (b) shows symmetric mode splitting in the reflection. (c) Demonstration of the non-symmetric splitting change with (*i*) two peaks' offsets and (*ii*) two peaks' height difference. The concept of applying non-symmetric splitting as circular scale is illustrated in the inset. (d) Comparison to the case without non-symmetric splitting, where the EP can only act as the main scale for measurement.

# Summary and outlook

We demonstrate a new type of non-Hermitian behavior in ring microresonators with multiple scatterers, arising from asymmetric coupling between intra-cavity modes and coupling waveguide modes. We find that under the right conditions, the forward propagating mode reaches maximal intensity near the exceptional point (EP), while the reflected mode is suppressed and exhibits a split resonance in its spectrum. By combining experimental measurements, finite element simulations and analytic modeling with a modified two mode approximation, we systematically characterize the asymmetric mode splitting in the reflection and the influence of the angular configuration of the two scatterers on the splitting behavior. The observed periodically repeating mode splitting patterns can be precisely engineered from using numerical simulations or an analytical model. This work provides new insight in non-Hermitian physics in fully-integrated optical systems on a chip-scale. We also perform a proof-of-concept analysis on a potential application of the non-symmetric mode splitting for EP-based sensing, which opens new pathways for sensitive detection of biomolecules, temperature fluctuations, displacements, refractive index changes or magnetic fields. Recent demonstrations of non-Hermitian symmetry breaking across different nanophotonic platforms, including acoustically-pumped time-reversal-symmetry breaking in resonators[36], magnetic-free on-chip optical isolation[37], Kerr-effect-induced symmetry breaking, optical switching in integrated microresonators[38,39], and spontaneous symmetry breaking in coupled nanocavities[40] highlight the wide applicability of non-Hermitian phenomena and motivate further exploration. By combining fundamental non-Hermitian physics with photonic chips, our work establishes a foundation for future applications in bio-sensing and integrated photonics.

**Acknowledgements**

The authors would like to acknowledge support from the Micro- and Nanostructuring department (MPL, Erlangen). This work has been supported by the European Union's H2020 ERC Starting Grant No. "CounterLight" 756966; Marie Curie Innovative Training Network "Microcombs" 812818, Max Planck Society, Max Planck School of Photonics, Max-Planck Fraunhofer Cooperation Project LAR3S, Munich Quantum Valley Project TeQSiC, and DFG Project 541267874.

# Supplemental Information: Scattering-Induced Mode Chirality in Ring Resonator

## I. Device fabrication and experimental setup

In this work, we fabricate our devices using commercial wafers with a 400-nm-thick silicon nitride layer and a 3-μm-thick silica base layer. Once the photonic structures are fabricated (by spin coating of photoresist, electron beam lithography, followed by developing and etching), a 2.5-μm-thick fused silica cladding is deposited by atomic layer deposition and plasma-enhanced chemical vapor deposition. Finally, thermal annealing is used to increase the $Q$-factor. The detailed experimental setup is shown in Fig. S1. In the experiment, light from the laser source is coupled into the chip via a lensed fiber. We use a microresonator in add-drop configuration to measure the transmission and reflection signals. One of the bus waveguides serves as the input port and we measure the transmitted and reflected signals from the second waveguide. Note that the input bus waveguide forms a low finesse Fabry-Perot cavity with the end facets of the chip, which would lead to unwanted inference effects if we would measure signals with the input waveguide. Thus, the add-drop configuration of the microresonator is beneficial for avoiding the unwanted interference. A polarization controller and a variable attenuator are used to control the polarization and the power of the input light. The output is connected to the photodiode. For both spectra, we initially maximize the transverse electric mode (TE) via the polarization controller. We tune the laser from 1510 nm to 1630 nm to measure the spectral response in both transmission and reflection. Part of the laser input (10 %) is sent to a reference fiber cavity, which is used to calibrate the frequency axis.

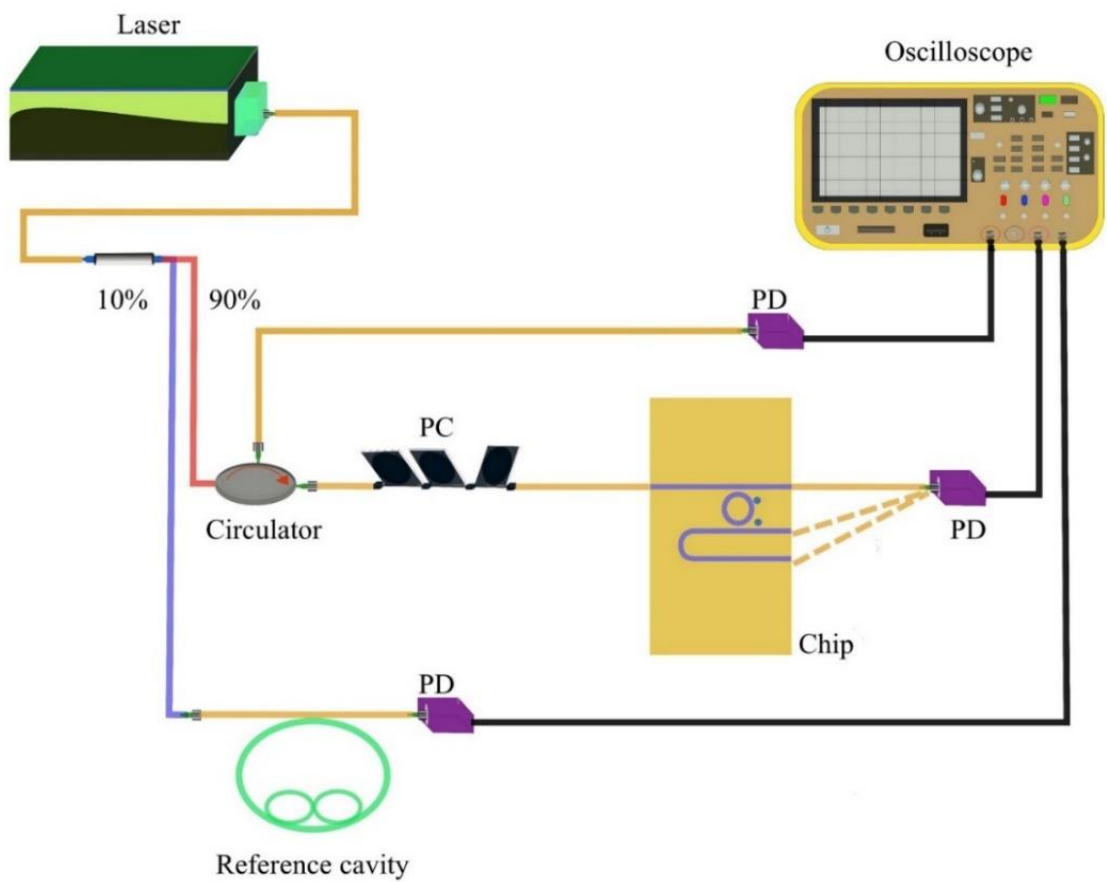

**Figure S1. Experimental setup.** A $Si_3N_4$ ring microresonator is pumped in one direction via a lensed fiber. A reference fiber cavity is used to calibrate the frequency offset between the cavity resonances. PC: polarization controller. PD: photodetector.

## II.  Mode spectroscopy

In this experiment we scan the laser to record the transmission and reflection spectra across a bandwidth of ~14 THz. In the reflection spectra, we observe a periodic behavior. In order to investigate this further, we fit the reflection spectra obtained from all the tested devices with a function $a \cdot e^{-b(x-c)^2} \cdot \sin^2(\omega f + \phi)$ where $a$, $b$, $c$ are constants, $f$ is the frequency from the spectrum, $\phi$ is the phase offset and $\omega$ is the period inside the gaussian envelope. Two selected experimental results and their corresponding fit are shown in Fig. S2, where (a) and (b) are corresponding to devices 2 and 3, for which the separation angle between the two scatterers can be found in the main text Fig. 3. By checking all the fits, we find that the period of the sinusoidal modulation $\omega$ varies in a small range from 2 to 2.3. For the demonstrated results in Fig. S2 specifically, the sinusoidal modulation $\omega$ has a value of 2.277 and 2.296 for (a) and (b) correspondingly. Such a periodic behavior in the reflection spectra may originate from interference effects introduced by the two scatterers in the ring resonator, and the modulation could also be attributed to the coupling between localized modes or wave interference within the separation between scatterers. In addition, we find that the reflection spectrum is very sensitive to the polarization, which indicates potential polarization-dependent scattering effects within the resonator.

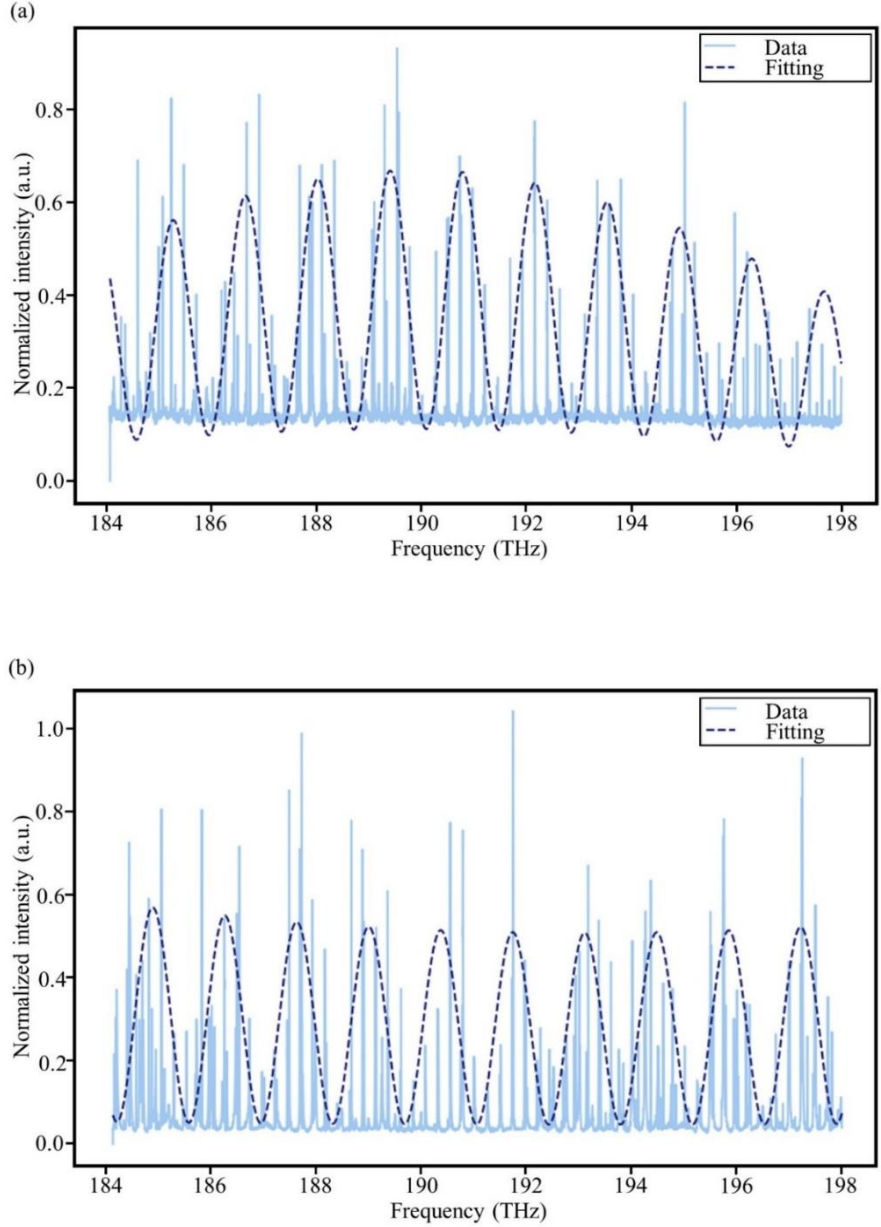

**Figure S2. Measured reflection spectra.** The spectra of the reflection signals with a fit using a $\sin^2$ function. (a) Data from device 2. (b) Data from device 3.

### III. Microresonators with scatterers

The numerical simulations in the main text are based on the model presented in Fig. S3(a), conducted using the "Electromagnetic Waves, Frequency Domain" module in COMSOL Multiphysics. For simplicity, we employ a two-dimensional model here. Relative to the experimental devices, the simulated structure is downscaled to reduce computation time, with parameters set as: $R = 50$ μm, $r = 0.25$ μm, $w = 1$ μm, $g = 0.5$ μm, and $d = 1$ μm. The cladding refractive index is set to $n = 1.5$, while the structure itself is given the refractive index of silicon nitride ($n = 2$). The top waveguide utilizes numerical ports combined with boundary mode analysis to define the incident conditions. The bottom waveguide is configured as an unexcited numerical port to collect transmitted and reflected signals, with boundary mode analysis performed prior to excitation. Scatterers are modeled with the same refractive index as the silicon nitride waveguide and positioned 50 nm from the resonator edge. Fig. S3(b) shows the spatial distribution of the $E_x$ field at the resonance wavelength within the simulated structure. At resonance, the electric field exhibits a distinct in-plane mode. As shown in the inset, most of the incident energy is coupled into the resonator, where forward and backward scattering occur near the scatterers. The energy flow at the output waveguides reveals both transmitted and reflected fields.

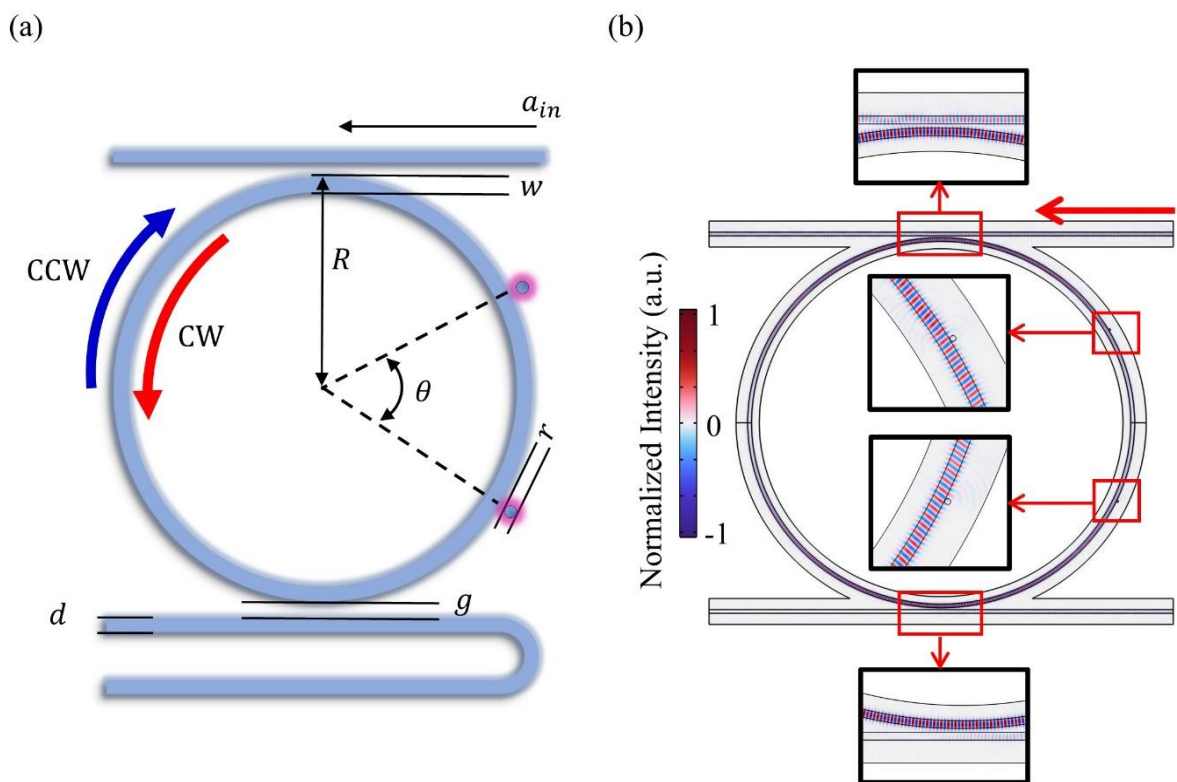

**Figure S3. Electric field distribution analysis.** We analyze the electric field distribution of our scatterer-disturbed ring microresonator system via a COMSOL simulation. We focus on the in-coupling and out-coupling regions as well as the regions near the scatterer as shown in the insets.

In the main text, we present detailed experimental results for different types of mode splitting (Fig. 2(a,b) summarize the splitting asymmetry across resonances). Here, we show the COMSOL simulation results for the transmission and reflection spectra across a relatively broad spectral range, as shown in Fig. S4. In this simulation, the separation angle between the two scatterers is fixed while the pump wavelength is continuously tuned from 1600 nm to 1640 nm. The resonances are shown in Fig. S4(a), where the transmission spectra and reflection spectra are shown together. Within this range, we observe nine resonances, with the reflection intensity consistently smaller than the transmission intensity. Fig. S4(b) shows three selected reflection splitting spectra, with their corresponding resonances labeled in Fig. S4(a). These three resonances represent nearly symmetric splitting, "left-low" asymmetric splitting, and "right-low" asymmetric splitting, all of which are also observed in our experiments. In each case, the reflection intensity is suppressed, and at the resonance wavelength the reflection can be further reduced, approaching nearly zero, consistent with both the experimental data and the simulations shown in the main text. These results confirm that our simulation analysis agrees with the experiment: it is capable of not only confirming the transmission and reflection spectra at one specific frequency, but also reproducing the patterns across wide pump detuning ranges if the physical parameters such as the separation angle between two scatterers, scatterer size, and scatterer-resonator distance are fixed. In addition, the pump detuning simulations are helpful to guide device design and experiments, as they allow us to predict which pump wavelength is most preferrable for a specific device, depending on the application requirements.

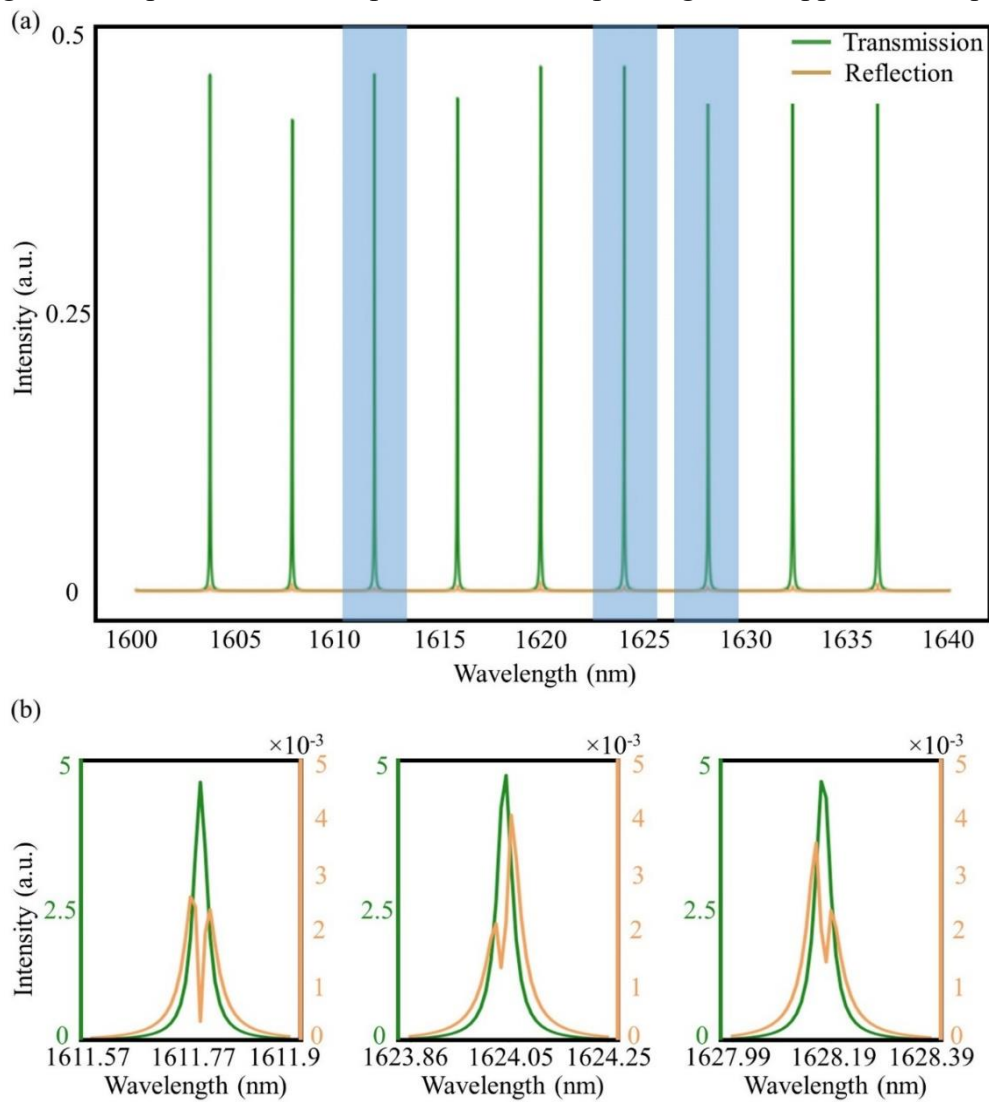

**Figure S4. Splitting simulation (COMSOL) at different resonances.** COMSOL simulations of splitting behavior across different resonances for a fixed angle between the two scatterers. (a) Transmission and reflection spectra over a ~40 nm span. The reflection intensities are much smaller than the transmission intensities. (b) Selected transmission resonances and reflection signals. The enlarged reflection spectra are shown below the selected transmission and reflection spectra. For the detected splitting peaks, asymmetric splitting dominates, and the degree of asymmetry shifts with the pump wavelength.

## IV. Intra-cavity mode analysis

In the main text, we provide a theoretical explanation for the origin of asymmetric mode splitting, which arises from the non-Hermitian coupling between intra-cavity modes and waveguide modes. Here, we extend the analysis of the shape of the intra-cavity modes. Starting from the governing equations presented in the main text, and assuming a constant input, we can drive the system into a stationary state within the cavity. In this steady-state regime, the complex mode amplitudes become time-independent, allowing the system's response function to be expressed analytically. In this system, both the transmission and reflection spectra are expected to display a characteristic "Lorentzian dip", as the response function takes the form:

$$\chi_{\text{ccw}} = \frac{2i\kappa \cdot (i\gamma + 2\Delta - 2K \cdot e^{i\delta_k})}{4\cos^2(m(\beta_1-\beta_2)) \cdot K^2 \cdot e^{2i\delta_k} + (\gamma - 2i\Delta + 2iK \cdot e^{i\delta_k})^2}$$

$$\chi_{\text{cw}} = \frac{B}{4\cos^2(m(\beta_1-\beta_2)) \cdot K^2 \cdot e^{2i\delta_k} + (\gamma - 2i\Delta + 2iK \cdot e^{i\delta_k})^2}$$

with weak backscattering from the scatterers and a constant CCW drive. The eigenvalues of the coupled mode equation are $\omega = \sqrt{AB} - \Omega$, and the corresponding eigenvectors are $v_\pm = \left(\pm\sqrt{A}, \sqrt{B}\right)^T$. As discussed in the main text, the system reaches the EP when the CW mode is effectively suppressed, leaving the CCW mode dominant. To further verify this argument, we plot the evolution of eigenvalues as functions of $k_1$ and $\beta_2$, as shown in Fig. S5(a) and S5(b), which are the real and imaginary parts of the two eigenvalues, respectively. From the visualized results, we can identify the EP at the point marked with an arrow, where the two eigenvectors becoming collinear and their eigenvalues coalesce. The appearance of EPs is periodic in the phase difference between two scatterers, which can be visualized by projecting the evolution of eigenvalues at a fixed $k_1$, as shown in Fig. S5(c). For the results presented in the main text, we observe that in the vicinity of the EP, interference effects due to waveguide backscattering become significant when the angular position of the scatterers is such that the cavity mode in one direction (CW or reflection) is suppressed. Notably, the same eigenfrequency behavior can also be reproduced in the simulations. Using COMSOL's eigenfrequency solver, we plot the evolution of the CW and CCW eigenfrequencies as a function of the angular separation between the scatterers, as shown in Fig. S5(d). At certain scatterer angles, the two resonant modes coalesce, marking the emergence of an EP. In the presence of an active waveguide, this EP manifests as unidirectional emission, characterized by extremely strong transmission and negligible reflection, consistent with results shown in the main text. Moving away from this angle lifts the degeneracy as the scattering effect diminishes, leading to the reappearance of both reflection and transmission. This is the basis for EP sensing under symmetric conditions.

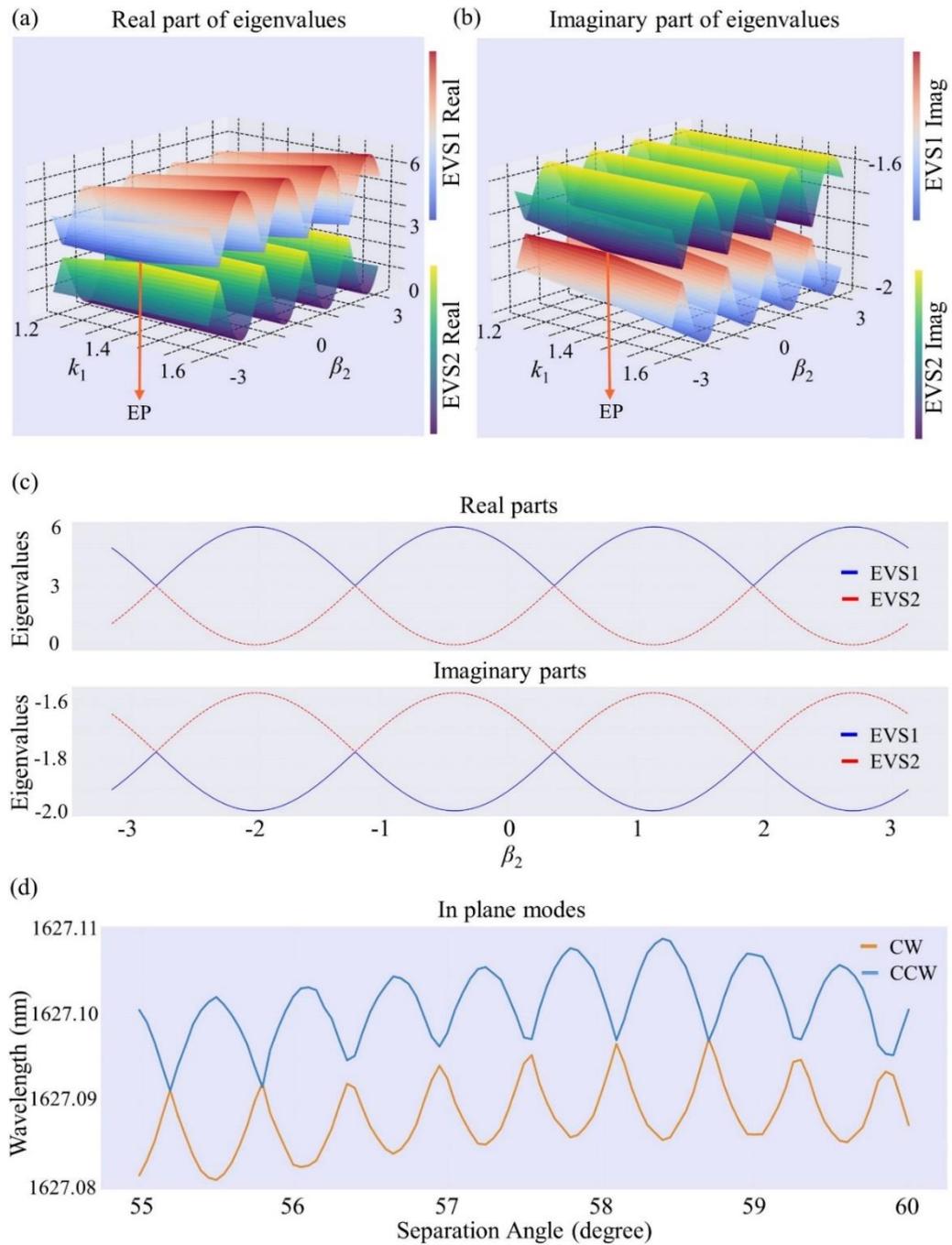

**Figure S5. Analysis of intra-cavity modes.** (a) and (b) Eigenvalue evolution as functions of phase angle for in-plane modes in the theoretical model, showing the locations of the exceptional points. (c) Projection of eigenvalues evolution at fixed $k_1$. (d) COMSOL simulation results for in-plane CW and CCW modes at different separation angles of the scatterers.

## V. Extended analysis of the mode splitting

To give a more comprehensive understanding of the splitting behavior and how the design can be optimized, we first vary the size of the microresonator and evaluate how the spatial dimension can affect the splitting behavior. All the results in this section are simulation results done by COMSOL. One selected result of splitting behavior change under continuous angle change between scatterers is shown in Fig. S6(a), corresponding to R = 30 μm. Since both the arclengths between the scatterers and the resonance frequency will change for microresonators with different sizes, we plot the splitting evolution in terms of multiples of wavelength. From Fig. S6(a) the pattern can be described as one non-split region, one split region and another non-split region when the scatterer separation increases. Although we only observe one split region for this case, the region here is larger than that shown in Fig. 4(b) which is corresponding to R = 50 μm. Note, the width of the resonance is also larger than the one shown in the main text Fig. 4(b), indicating that we have a lower Q-factor resonance at a similar pump wavelength. Consequently, the splitting width here is slightly larger compared to the R = 50 μm case, consistent with the trend observed in the main text Fig. 3(d). The summarized results for splitting width change in continuous scatterer separation angle change for five different microresonator sizes are shown in Fig. S6(b). From this plot we can see that both the average splitting width and splitting region width changes as the radius changes, and the average splitting width increases when the radius decreases. This relation can be intuitively attributed to the Q factor decrease as the radius decreases since the small rings will suffer from larger bending loss, which broadens the resonance and thereby increase the splitting width. These results intuitively guide us on how to engineer the desired splitting behavior in considering the balance between microresonator size and the splitting width and splitting region span.

In the main text, we analyze the case where one scatterer is fixed and the second one is moved. From the theoretical analysis, the coupling between the modes from the waveguide and ring contributes to the formation of the asymmetric splitting. To further investigate, we fix the angular separation between two scatterers and make them move together, and we consider the analysis here all for R = 50 μm. We first fix the separation angle between the two scatterers at 55 degree, which is a separation at which we can observe splitting during the continuous angle change of only one single scatterer. The splitting behavior change is plotted as a function of detuning and arclength in terms of multiples of wavelength. Here the arclength is the distance that two scatterers move around the ring while the separation between these two scatterers is fixed. The result is shown in Fig. S7(a), corresponding to the same resonance as in the main text. Here the observed pattern, that the splitting appears and disappears much more frequently, is distinct from the case in the main text. One cannot clearly identify regions for splitting or non-splitting. As shown in Fig. S7(b) and Fig. S7(c), both the splitting peak frequency asymmetry and height asymmetry are highly sensitive to the change of arclength across the entire region, i.e., sensitive to the position of the two scatterers, and the behavior is also not strictly periodic. For example, as shown in Fig. S7(b), under synchronous motion of the two scatterers, while resonance peak shift demonstrate periodicity, the resulting spectral splitting is non-repetitive. The splitting behavior is characterized by the frequency asymmetry plus the splitting width asymmetry, this phenomenon is more clearly demonstrated in Fig. S7(c), we assume that the splitting peaks never exactly repeat even though the resonance peak shifts are much more periodic here. These results demonstrate that that beyond the regular splitting discussed in the main text Fig. 4, it is possible to create finer-step sensitive splitting, although its emergence and disappearance is non-repetitive, which has a relatively clearly-defined region, the successive appearance and disappearance of the splitting at small steps is also feasible and may be useful for specific applications.

Another example is shown in Fig. S8, where we change the angle separation between the two scatterers from 55 degree to 57.3 degree, which is corresponding to the "no splitting"-case and is observed during the continuous scatterer angle change in the main text. Interestingly though, this specific angle separation does not yield any splitting, however, we observe a splitting pattern as shown in Fig. S8(a) when the two scatterers move together. The quantified peak frequency asymmetry and height asymmetry are shown in Fig. S8(b). The results here demonstrate new features, where we observe a dominant splitting region, which resembles the Fig. 4 results, plus a more frequent appearance and disappearance of splitting outside this regime, which resembles the Fig. S7 results. Again, no clear periodicity can be found here. Notably, the results in Fig. S8 provide another new perspective of engineering the patterns. Synchronous motion at a fixed scattering angle provides a phase-like effect that determines the splitting behavior at the initial angle, but this does not affect the periodic splitting behavior induced by varying the scatterer angle as described in the main text. Given that the splitting emerges concurrently with the suppression of reflection, the introduction of an initial phase can be adapted to more complex application scenarios.

Taking together the results shown in Fig. S7 and S8, as well as the results in Fig. 4, we demonstrate how one can design the desired splitting behavior via simulations. In addition, the results here further verify that the non-Hermitian coupling between the cavity modes and waveguide modes will be influenced by the position of both scatterers. From the perspective of engineering, to obtain the desired splitting behavior with large scatterer separation angle changes, a parameter list including initial position of the scatterers, relative position of the scatterers, distance between scatterers and ring edge, etc. can be considered. Other parameters like scatterer size, morphology, numbers of the scatterers, and their combinations with the current parameter list can also be explored. Another direction inspired by these results would be a mapping between the effective parameters of the theoretical model and the physical parameters in the simulation model.

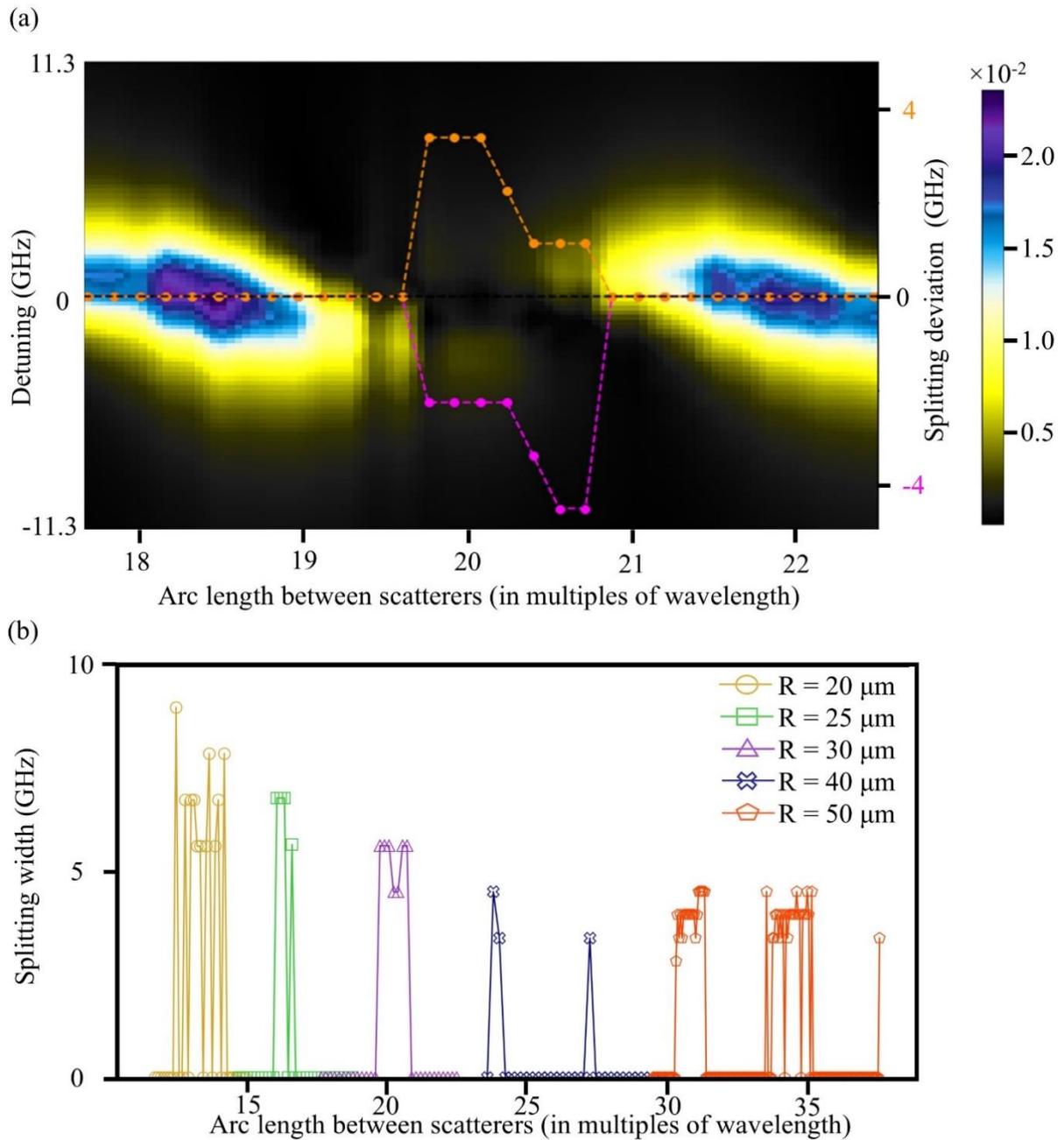

**Figure S6. Analysis of the mode splitting behavior under microresonator size variation.** We vary the radius of the microresonator and analyze the change of splitting behavior in a COMSOL simulation. (a) Simulation result of the reflection splitting evolution under continuous scatterer angle variation for a microresonator with R = 30 μm, shown as a 2D heatmap. (b) Summary of splitting width variation under continuous scatterer angle variation for different microresonators whose radius are ranging from 20 microns to 50 microns. The scatterer separation angle is shown as arc length in multiples of wavelength.

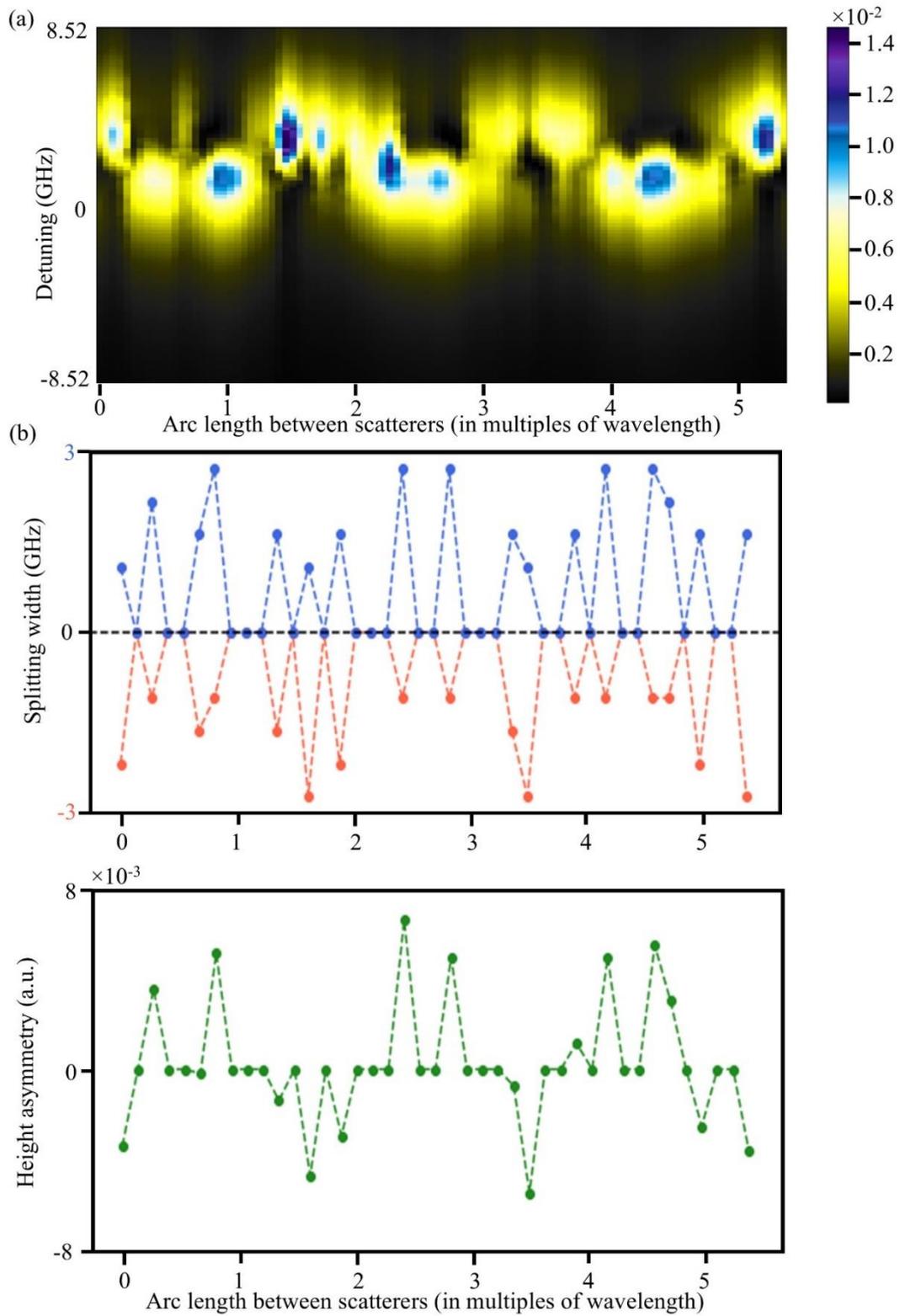

**Figure S7. Analysis of the splitting behavior when moving the two scatterers together at a fixed separation angle of 55 degrees.** (a) Simulation result of the reflection for a microresonator with R = 50 microns. (b) Summary of the splitting width and height asymmetry evolution.

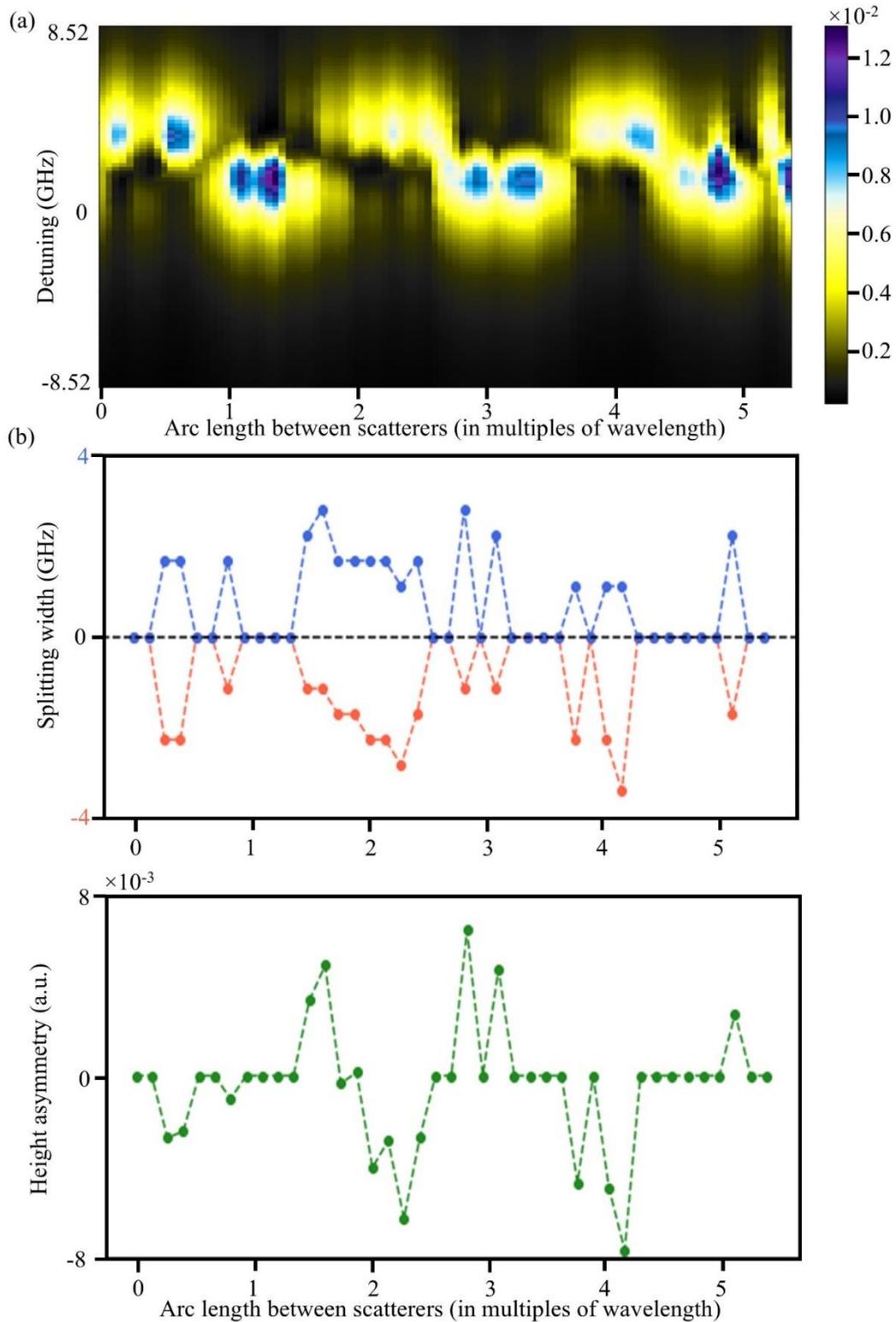

**Figure S8. Analysis of splitting behavior when moving the two scatterers together at a fixed separation angle of 57.3 degrees.** (a) Simulation result of the reflection for a microresonator with R = 50 microns. (b) Summary of the splitting width evolution and height asymmetry evolution.

## VI. Extended theoretical model and simulation results of the splitting evolution when continuously changing the scatterer angle

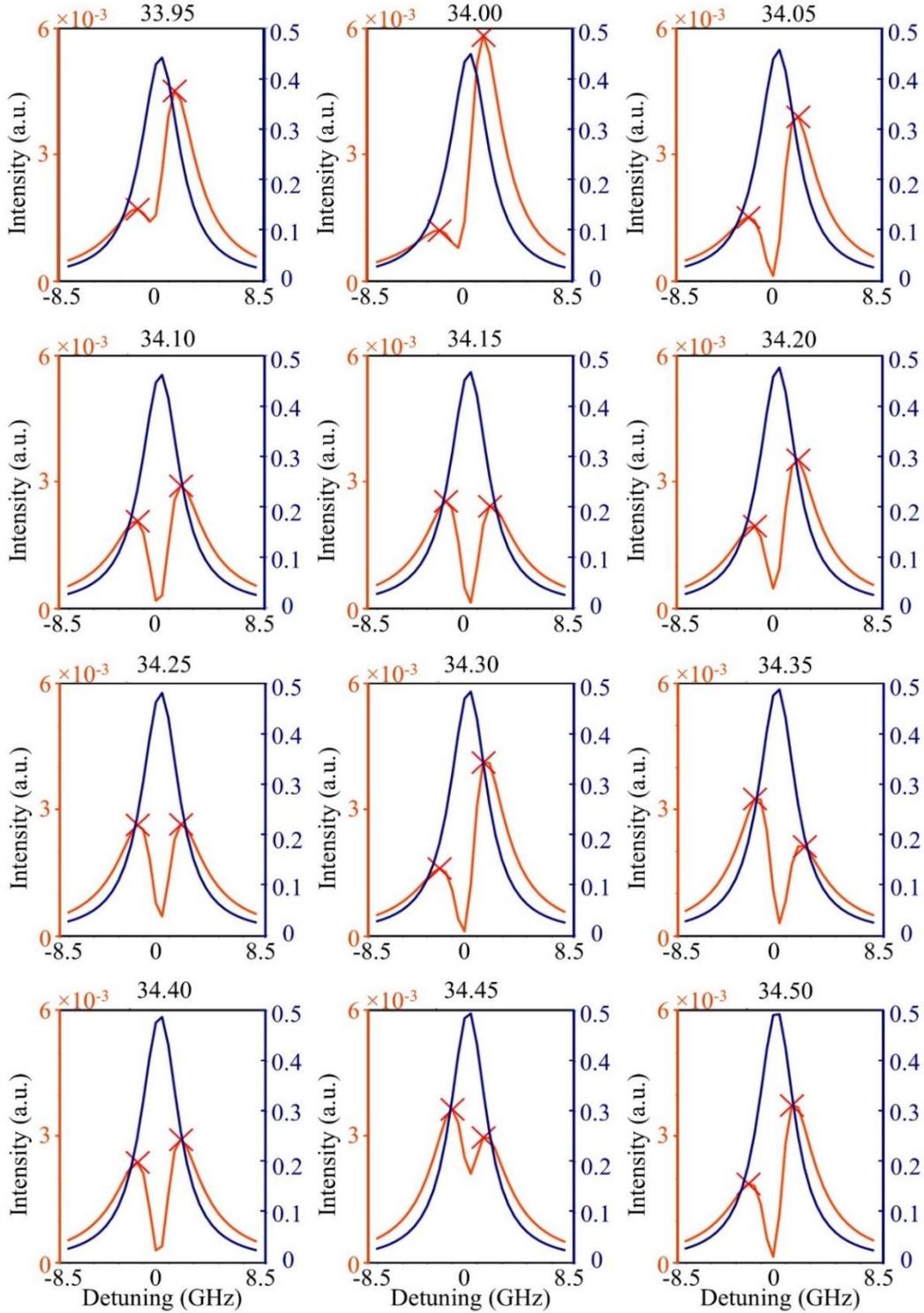

**Figure S9. Extended simulation data for continuous changes of the scatterer angle.** A detailed look at the splitting evolution in small steps of 0.05 multiples of the wavelength. The corresponding normalized angles are shown as title in each graph.

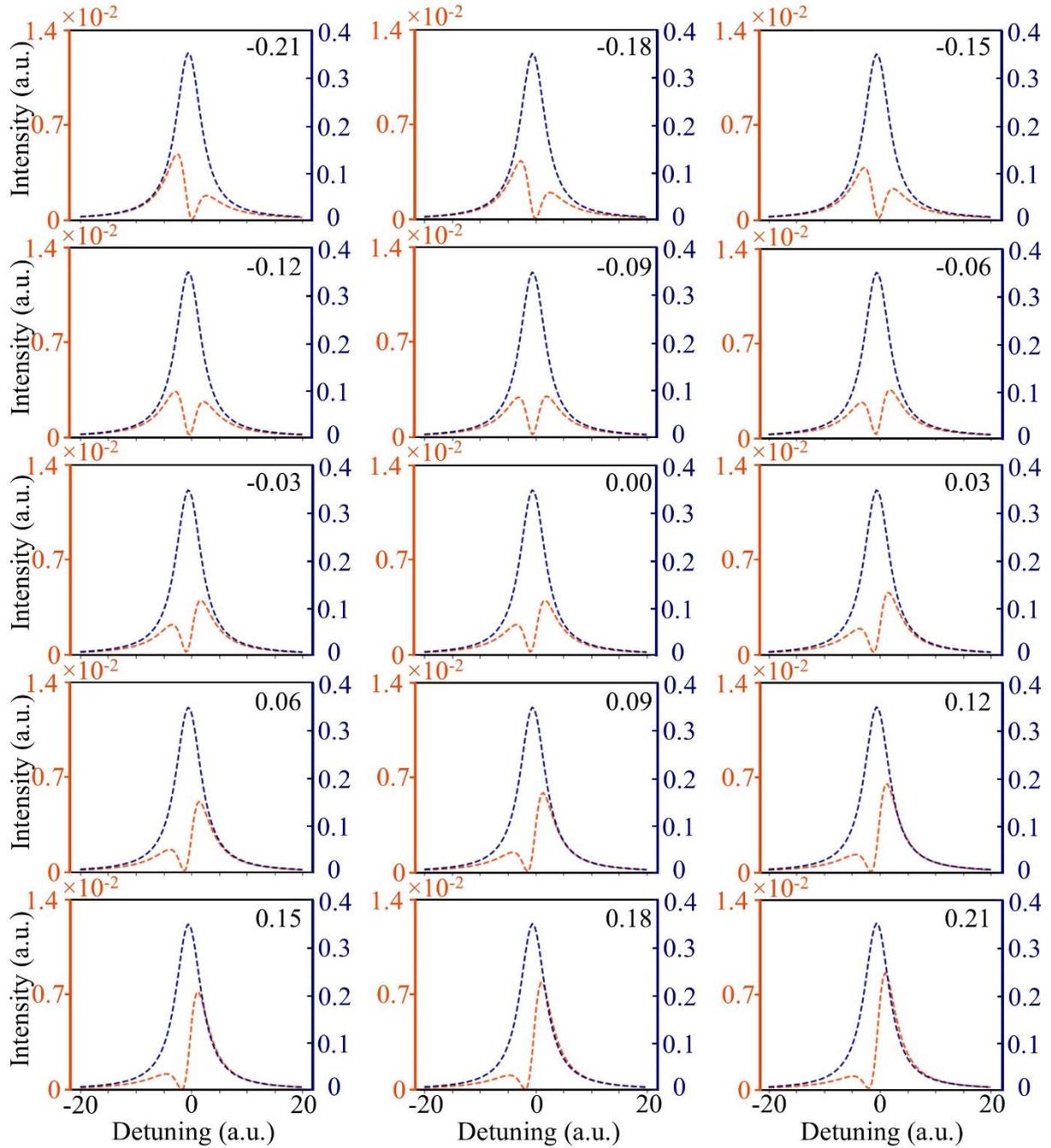

**Figure S10. Extended theoretical data for the continuous scatterer angle variation.** A detailed look at the splitting evolution in small steps of the scatterer separation angle. We observe a complete shift from left-dominant asymmetry to right-dominant asymmetry. The scatterer separation in radians is shown in the top-left of each panel.